\documentstyle[prc,aps]{revtex}
\begin{document}
\draft
\title {Ground-state energies and widths of $^5$He and $^5$Li nuclei}
\author{V.D.~Efros\thanks{Permanent address: Russian Research Centre
"Kurchatov Institute", 123182 Moscow, Russia}
and H.~Oberhummer}
\address{Institut f\"ur Kernphysik, Wiedner Hauptstr.~8--10, TU Wien,
A--1040 Vienna, Austria}
\date{\today}
\maketitle
\begin{abstract}
We extract energies and widths of the ground states
of $^5$He and $^5$Li
from recent single--level R--matrix fits to the
spectra of the $^3$H$({\rm d},\gamma$)$^5$He and
the $^3$He$({\rm d},\gamma$)$^5$Li reactions.
The widths obtained differ significantly from the formal
R--matrix values but they are close to those measured as
full widths at half maxima
of the spectra in various experiments. The energies are somewhat
lower than those given by usual estimates of
the peak positions. The extracted values are close to
the S--matrix poles calculated previously from the multi--term analyses of the
N-$^4$He elastic scattering data.
\end{abstract}



The ground--state properties of $^5$He and $^5$Li nuclei were studied
in several experiments, see Refs.~\cite{ajz88,lau66,bal91} for a
review.  In the most recent work~\cite{bal91} the values of the
$A=5$ widths were reported which are significantly larger than
the previous ones.  In the present work we reconsider the
experimental data of Ref.~\cite{bal91}. 
We extract the widths and the positions of
the {A=5} levels from these data. 

The present work has been done in connection with some
applications of an astrophysical interest \cite{gor95,ohu95,efr95}. Besides,
one may note that at present there exist difficulties in obtaining
p--wave N-$^4$He phase shifts at low--energy from a realistic non--central
NN interaction (see e.g. \cite{ust89}) which hinders a microscopic 
description of light multicluster nuclear systems (see e.g. \cite{cso93}).
The phase shifts are governed by the $^5$He and $^5$Li ground--state
resonances and it would be convenient to analyze the problem directly in
terms of their energies and widths. 
 
In Ref.~\cite{bal91} a single--term R--matrix fit to experimental
$\gamma$--ray spectra of that work was undertaken to determine the widths,
while in most
of the previous work the widths were estimated by measuring FWHM of the
spectra (see Table 1 in Ref.~\cite{bal91}). The
ground--state energies were mainly estimated from the positions of the peaks
(see~\cite{lau66}). In Ref.~\cite{bond77} the $^5$He energy and width were
extracted from an R--matrix fit to the whole set of the
$^4$He(n,n)$^4$He cross section and polarization data in the elastic region.
The p--wave R--matrix elements were approximated by resonant terms
plus linear functions in energy representing contributions from distant
R--matrix levels. The position of the pole of the expression obtained was
found giving the energy and width of $^5$He. In Ref.~\cite{kraus74} a
similar
type analysis was carried out for $^4$He(n,n)$^4$He and $^4$He(p,p)$^4$He
scattering using another form of the S--matrix parameterization. These
multi--term analyses are considerably influenced by potential scattering
interference. It is of interest to
study the pole structure of the S--matrix for other types of processes
involving $A=5$ nuclei.

In Ref.~\cite{bal91} the $^3$H(d,$\gamma)^5$He and
the $^3$He(d,$\gamma)^5$Li spectra have been
measured at $E_{\rm d}=8.6$\,MeV and
$\theta_{\rm lab}=90^{\rm o}$. The spectra were
fitted by the expression\footnote{In Eq.~(1) we
express the spectra in terms of
the energy
$E_{{\rm N}\alpha}$ that coincides
with $E_{\rm N}$ from
Ref.~\cite{bal91}. Our $E_{\rm res}$
coincides with that of
Ref.~\cite{bal91} in their Table II but differs from $E_{\rm res}$
in their Eq.~(1).
Our quantity $\Gamma_R(E)$ coincides with $\Gamma_{\rm N}(E)$
from Ref.~\cite{bal91}.}
\begin{equation}
\frac{\Gamma_R(E_{{\rm N}\alpha})}{[E_{{\rm N}\alpha}-E_{\rm
res}-\Delta(E_{{\rm N}\alpha})]^2+
\left[\frac{\Gamma_R(E_{{\rm N}\alpha})}{2}\right]^2} \label{eq:rmatr}
\end{equation}
times a slowly varying function in the energy $E_{{\rm N}\alpha}$. Here
$E_{{\rm N}\alpha}$ is the relative motion energy of the (N+$\alpha$)
subsystem, and
$\Delta(E_{\rm res})=0$.  The energy dependencies of the width
$\Gamma_R(E)$ and of the shift function $\Delta(E)$ corresponded to the
single--level R--matrix expression, that is
$\Gamma_R(E)=2\gamma^2P(E)$ and
$\Delta(E)=-\gamma^2[S(E)-S(E_{\rm res})]$.
Here $\gamma^2$ is the reduced width, while
the penetrability $P$ and the shift function $S$
for the $\ell=1$ orbital
momentum are defined as usual ~\cite{lan58}. The overall normalizations of
the spectra and
the reduced widths were fitted to the experimental data while for the
$E_{\rm res}$ energies
the
values from Ref.~\cite{ajz88} were used. As a result the following values
of the widths
$\Gamma_R(E_{\rm res})$ were obtained which we shall denote $\Gamma_R$
\begin{equation}
\Gamma_R(^5\mbox{He})=(1.36\pm 0.19)\,\mbox{MeV},\,\,\,\,\,\,\
\Gamma_R(^5\mbox{Li})=(2.44\pm 0.21)\,\mbox{MeV}.  \label{eq:rwidths}
\end{equation}
Here and below the input parameters
are \cite{bal91}: the channel radius
$R=3.6$\,fm, $E_{\rm res}=0.89$\,MeV,
$\gamma^2=(3.32\pm 0.46)$\,MeV in the $^5$He case,
and $R=3.6$\,fm, $E_{\rm res}=1.97$\,MeV,
$\gamma^2=(3.33\pm 0.29)$\,MeV in the $^5$Li case.
In Eq.~(\ref{eq:rwidths}) and below the errors correspond to those
in the reduced widths.

The previous
values of the width ranged
from $(0.55\pm 0.03)$\,MeV to $(0.85\pm 0.05)$\,MeV in the $^5$He case.
In the $^5$Li case the previous
widths were grouped
around 1.5\,MeV except for one value of $(2.6\pm 0.4)$\,MeV. The latter
value \cite{bus68}
was obtained using the method in principle similar to that of Ref.
\cite{bal91}, while
the others were
obtained in several experiments by
measuring FHWMs of spectra for various reactions
(see~\cite{ajz88,bal91}) and with the help of the above--mentioned
analyses~\cite{bond77,kraus74} of the elastic N--$^4$He scattering. The
FWHM for
the spectra of Ref.~\cite{bal91} are rather close to the above--mentioned
values.

We shall extract the poles $E=E_0-i\Gamma/2$
of the scattering matrix from the results of
Ref. \cite{bal91}. These poles determine
the {\em physical} resonance energies $E_0$ and widths $\Gamma$ that
are
different from  $E_{\rm res}$ and $\Gamma_R=\Gamma_R(E_{\rm res})$
entering Eq.~(\ref{eq:rmatr}). The difference proves sometimes to be quite
significant,
see e.g.~\cite{hal87}.
In particular, namely $\Gamma$ and not
$\Gamma_R$ determines the lifetime of a system as it is
required for example in the applications~\cite{gor95,ohu95,efr95}.
One can see that $\Gamma_R(E_{\rm res})$ cannot serve as the estimate of
a width even in a narrow resonance case.  The quantity
$\Gamma_S(E)=[1-\Delta'(E_{\rm res})]^{-1}\Gamma_R(E)$ at $E=E_{\rm res}$
could have served as such an estimate \cite{lan58}. Here
$\Delta'$ is a derivative over $E$ taken at the point $E_{\rm res}$.
(The quantity $\Gamma_S(E)$ enters  the resonant factor of a reaction
when written in the form $\Gamma_S(E)/[(E-E_{\rm
res})^2+(\Gamma_S(E)/2)^2]$.) In our case the widths are broad and the
above--mentioned estimate is not very accurate, however.

In case of the parameterization of Eq.~(\ref{eq:rmatr}) the complex
energy $E$ of the resonance is the solution to the equation
\begin{equation}
[E-E_{\rm res}-\Delta(E)]^2+[\Gamma_R(E)/2]^2=0, \label{eq:pole1}
\end{equation}
at the condition $k=(2\mu E)^{1/2}/\hbar=k_1-ik_2$, $k_1>0,k_2>0$.
In the $^5$He case this equation
can be represented in the form
\[ (x-x_{\rm res})^2[\gamma^{-2}\bar{E}\cdot (x+1)+(x_{\rm
res}+1)^{-1}]^2+x^3=0, \]
where we introduced the notation $\bar{E}=\hbar^2/(2\mu R^2)$, $\mu$ being the
reduced mass, and $x=E/\bar{E}$, $x_{\rm res}=E_{\rm res}/\bar{E}$.
We obtained\footnote{A pleasant peculiarity of the result of
Eq.~(\ref{eq:widths2}) for the
width is
a considerable reduction of the relative errors as compared to
Eq.~(\ref{eq:rwidths}). The same holds true for
Eqs.~(\ref{eq:widths3})--(\ref{eq:widths5}) below and this is due to the
difference
in the $\gamma^2$--dependence in the equations determining the widths. Note
as well
that the highest (lowest) $E_0$ value, of course,
corresponds to the lowest (highest) $\Gamma$
value in Eq.~(\ref{eq:widths2}) and in the relations below.}
\begin{equation}
E_0(^5\mbox{He})=(0.735\pm 0.02)\,\,\mbox{MeV},\,\,\,\,\,\,\,
\Gamma(^5\mbox{He})=(0.57\pm 0.02)\,\,\mbox{MeV}. \label{eq:widths2}
\end{equation}
Note that the value of $E_0$ obtained is slightly shifted downwards with
respect to the position of the peak of the spectrum which position equals to
0.81\,MeV. This is due to an asymmetric form of the spectrum.

The values (\ref{eq:widths2}) should be somewhat corrected due to the
following reason. The value of 0.89\,MeV taken in Ref. \cite{bal91} as
$E_{\rm res}$ was actually obtained \cite{lau66} as the mean
position of the resonant peak. In general, this position differs from
the optimal $E_{\rm res}$ value or, which is the same, from the energy at
which the
resonant phase shift reaches $90^{\rm o}$.
We changed the value of $E_{\rm res}$
to shift the position of the peak to the value of 0.89\,MeV. The resulting
$E_{\rm res}$ value, $E_{\rm
res}=0.98$\,MeV, is close to those given in Refs.
\cite{ajz88,bond77,sta72} for the elastic $n-^4$He scattering. It is not
clear whether the small change in the $E_{\rm res}$ value made is quite
compatible with the $\gamma^2$ value deduced in Ref. \cite{bal91} from
the fit to the data.  The point is that in order to allow for spectrum
calibration errors the fitting procedure in Ref.  \cite{bal91} included
shifts up or down in energy of the entire observed spectrum
resulting from a convolution with a detector response function.
The information on the latter function listed in the paper \cite{bal91}
does not suffice
to perform the convolution.
If, nevertheless, one adopts the shifted $E_{\rm
res}$ value one obtains the following energy and width of $^5$He:
\begin{equation}
E_0(^5\mbox{He})=(0.80\pm 0.02)\,\mbox{MeV},\,\,\,\,\,\,\,
\Gamma(^5\mbox{He})=(0.65\pm 0.02)\,\mbox{MeV}. \label{eq:widths3}
\end{equation}
These values
are in a remarkable agreement with the values of $E_0=0.77$\,MeV and
$\Gamma=0.64$\,MeV, and also with the values of $E_0=0.78$\,MeV and
$\Gamma=0.72$\,MeV obtained in Refs.~\cite{bond77} and \cite{kraus74},
respectively, from multi--term analyses of the elastic n--$^4$He
scattering. The measured FWHM (see Table 1 in Ref.\cite{bal91}) are
also in accordance with the width value obtained.

In the $^5$Li case, $\Delta(E)$ and $\Gamma_R(E)$ entering
Eq.(\ref{eq:pole1}) include Coulomb functions with complex arguments
and their derivatives. They were calculated with the help of the
computer code of Ref.~\cite{tho85}. Eq.~(\ref{eq:pole1}) was solved
with a Newton--type method.  The results corresponding to
Eqs.~(\ref{eq:widths2}) and (\ref{eq:widths3}) are
\begin{equation}
E_0(^5\mbox{Li})=(1.635\pm 0.03)\,\mbox{MeV},\,\,\,\,\,\,\,
\Gamma(^5\mbox{Li})=(1.16\pm 0.03)\,\mbox{MeV}, \label{eq:widths4}
\end{equation}
and
\begin{equation}
E_0(^5\mbox{Li})=(1.72\pm 0.03)\,\mbox{MeV},\,\,\,\,\,\,\,
\Gamma(^5\mbox{Li})=(1.28\pm 0.03)\,\mbox{MeV}, \label{eq:widths5}
\end{equation}
respectively.
The $E_0$ values are in accordance with the value of $E_0=1.6$\, MeV
obtained in Ref. \cite{kraus74} from the multi--term analysis of the elastic
p--$^4$He scattering. Similar to the $^5$He case these values are
shifted downwards with respect to the peak position of 1.97\,MeV. The
widths obtained are lower than the value of $\Gamma=1.45$\,MeV from
Ref. \cite{kraus74} and than the mean FWHM value \cite{ajz88} of
$\approx 1.5$\,MeV, but
they are close to the FWHM value of $(1.24\pm 0.03)$\,MeV for the spectrum
of Ref.
\cite{bal91}.

It may be noted that the above $S$--matrix pole widths cannot be
"reduced" as were the $R$--matrix widths in Ref. \cite{bal91} to obtain
reduced widths $\gamma^2$ . The reduced widths of $^5$He and $^5$Li
extracted
in Ref. \cite{bal91} turned out to be equal which showed that 
charge symmetry is not violated. The ratio of the $R$--matrix widths
\cite{bal91} thus has 
a direct physical meaning as well as the $S$--matrix pole results.

In conclusion, the experimental data of Ref.~\cite{bal91} on the
$^3$H$(d,\gamma$)$^5$He and the $^3$He$(d,\gamma$)$^5$Li
reactions allowed safe extrapolations to the S--matrix
poles. From these extrapolations
rather accurate values of the complex energies of $A=5$ nuclei have
been determined.
They are in a good agreement with those obtained from
the multi--term analyses of N--$^4$He cross section and polarization
data set in the elastic region. In spite of the broad
resonances the widths obtained proved to be close to those
measured as FWHM.  However, the real parts of the poles are shifted
downwards with respect to the positions of the peaks in the spectra. 

This work was supported
by Fonds zur F\"orderung der wissenschaftlichen
For\-schung in \"Osterreich
and by the International Science Foundation and Russian Government
(grant J4M100).


\begin{references}
%
\bibitem{ajz88} F.~Ajzenberg-Selove and T. Lauritsen, Nucl.~Phys.~{\bf A227}, 1
(1974);
F.~Ajzenberg-Selove, Nucl.~Phys.~{\bf A320}, 1 (1979);
{\bf A413}, 1 (1984); {\bf A490}, 1 (1988).
%
\bibitem{lau66} T. Lauritsen and F. Ajzenberg-Selove, Nucl.~Phys.~{\bf 78},
1 (1966).
%
\bibitem{bal91} M.J.~Balbes, G.~Feldman, L.H.~Kramer, H.R.~Weller,
and D.R.~Tilley, Phys.~Rev.~C {\bf 43}, 343 (1991).
%
\bibitem{gor95} J.~G\"orres, H.~Herndl, I.J.~Thompson, and M.~Wiescher,
Phys.~Rev.~C {\bf 52}, 2231 (1995).
%
\bibitem{ohu95} H.~Oberhummer, W.~Balogh, V.D.~Efros, H.~Herndl,
and R.~Hofinger, Few--Body Sys. Suppl., {\bf 8}, 317 (1995).
%
\bibitem{efr95} V.D.~Efros, W.~Balogh, H.~Herndl, R.~Hofinger,
and H.~Oberhummer, Z.~Physik A, 1996, in press.
%
\bibitem{ust89} M.N.~Ustinin and V.D.~\'Efros, Yad.~Fiz. {\bf 49}, 1297
(1989) [Sov.~J.~Nucl.~Phys. {\bf 49}, 807 (1989)].
%
\bibitem{cso93} A. Cs\'ot\'o, Phys.~Rev.~C {\bf 48}, 165 (1993).
%
\bibitem{bond77} J.E. Bond and F.W.K. Kirk, Nucl.~Phys.~{\bf A287}, 317 (1977).
%
\bibitem{kraus74} L. Kraus and L. Linck, Nucl.~Phys.~{\bf A224}, 45 (1974).
%
\bibitem{lan58} A.M. Lane and R.G. Thomas, Rev.~Mod.~Phys.~{\bf 30}, 257 (1958).
%
\bibitem{bus68} W. Buss, W. Del Bianco, H. Waffler, and B. Ziegler,
Nucl.~Phys.~{\bf A112},
47 (1968).
%
\bibitem{hal87} G.M. Hale, R.E. Brown and N. Jarmie, Phys.~Rev.~Lett.~{\bf 59},
763 (1987).
%
\bibitem{sta72} Th. Stammbach and R.L. Walter, Nucl.~ Phys.~{\bf A180},
225 (1972).
%
\bibitem{tho85} I.J. Thompson and A.R. Barnett, Comput.~Phys.~Comm.~{\bf
36}, 363
(1985).
%
\end{references}
\end{document}